\documentstyle[12pt]{article}
 \newfont{\goth}{cmbx9}
 \newfont{\gothi}{cmbx9}

\newtheorem{defi}{Definition}
\newtheorem{theo}{Theorem}
\newtheorem{prop}{Proposition}
\newtheorem{lemma}{Lemma}
\newtheorem{rem}{Remark}
\newcommand{\bdef}{\begin{defi}}
\newcommand{\ede}{\end{defi}}
\newcommand{\bsat}{\begin{theo}}
\newcommand{\esat}{\end{theo}}
\newcommand{\bprop}{\begin{prop}}
\newcommand{\eprop}{\end{prop}}
\newcommand{\blem}{\begin{lemma}}
\newcommand{\elem}{\end{lemma}}
\newcommand{\brem}{\begin{rem}}
\newcommand{\erem}{\end{rem}}

\newcommand{\be}{\begin{equation}}
\newcommand{\ee}{\end{equation}}

\newcommand{\ra}{\rightarrow}

\newcommand{\f}{\frac}

\begin{document}
\title{Maslov indices for periodic orbits\\~
       \\
       Contribution presented at the XIX ICGTMP Salamanca}

\author{E. Meinrenken}

\date{\normalsize
       Fakult\"at f\"ur Physik der Universit\"at Freiburg, \\
       Hermann-Herder-Str.~3, D-7800 Freiburg / FRG }

\maketitle

\thispagestyle{empty}

\begin{abstract}
\noindent

It is shown that there is a generalization of the Conley-Zehnder index
for periodic trajectories of a classical Hamiltonian system $(Q,\omega,H)$
from the
case $Q=T^*{\bf R}^n$ to arbitrary symplectic manifolds.
As it turns out, it is precisely this index which appears as a `Maslov phase'
in the trace formulas by Gutzwiller and Duistermaat-Guillemin.
\newline
\end{abstract}

\vfill
 \begin{flushright}
 \parbox{12em}
 { \begin{center}
 University of Freiburg \\
 THEP 92/21 \\
 September 1992
 \end{center} }
 \end{flushright}

\newpage
\renewcommand{\thefootnote}{\arabic{footnote}}
\setcounter{page}{1}

\noindent {\Large\bf{Maslov indices for periodic orbits}}  \newline
\begin{center}
{\large {\bf E. Meinrenken}\\
 Universit\"at Freiburg, Fakult\"at f\"ur Physik,\\ Hermann-Herder Str.3,
 D-7800 Freiburg}
 \end{center}

\addtocounter{section}{-1}
\section{Introduction}

Consider a classical mechanical system $(Q,\omega,H)$, where $Q$ is a
symplectic manifold, $\omega$ its symplectic 2-form and
$H\in C^\infty(Q,{\bf R})$
a Hamiltonian. Assume that $\gamma:{\bf R}\ra Q$ is a periodic
solution curve of period $T$ and energy $E$ for the Hamiltonian vector field
$X_H$ on $Q$. The closed trajectory is called {\em regular} if the
(linear) Poincar\'e map $P(T)$ has no unit eigenvalue. From the implicit
function
theorem it follows that regular periodic trajectory always come in 1-parameter
families. Their `orbit cylinder' is a symplectic
submanifold of $Q$ which is transversal to all energy surfaces $H^{-1}(E)$.
We refer to \cite{AM78} for a detailed proof, which also shows that orbit
cylinders are stable under small perturbations of the Hamiltonian.

Conley and Zehnder \cite{CZ84} have defined
an index ${\rm ind}_{CZ}(\gamma)$ for
regular periodic orbits in $T^*{\bf R}^n$, generalizing the
usual Morse index for closed geodesics on a Riemannian manifold.
Roughly speaking, the index measures how often neighbouring
trajectories of the same energy wind round the orbit. It is stable under
deformations of the orbit as long as the regularity assumption is not
violated. In particular, all members of the orbit cylinder have
the same index.

As we will see below, the Conley-Zehnder index admits a natural extension
to arbitrary symplectic manifolds. The construction will only depend on the
choice of
a homotopy class of Lagrangian subbundles $L$ of $TQ$ along
the orbit $\gamma^\sharp=\gamma({\bf R})$. Such a choice is often dictated
by the particular system under
study, and is natural e.g. for cotangent bundles $Q=T^*X$ or if the orbit
is contractible. The index ${\rm ind}(\gamma,L)$
is characterized by
the following two properties:
\begin{itemize}
\item[I1] The index is stable under small perturbations of the system
      (as long as the orbit remains regular).
\item[I2] Whenever there is an invariant Lagrangian subbundle $M$ of $TQ$ along
      the orbit, ${\rm ind}(\gamma,L)$ is the (Maslov) intersection number
      of $M$ with $L$.
\end{itemize}

We show in \cite{M92} that it is precisely this index that appears as
a Maslov phase in the `trace formulas' by Gutzwiller \cite{G71} and
Duistermaat-Guillemin \cite{DG75}. For closely related results, see
Duistermaat \cite{D76} and Robbins \cite{R91}.

\section{The Intersection Number of Lagrangian Subspaces}

Let $(E,\omega)$ be a real symplectic vector space of dimension $2n$
and let $\Lambda(E)$ be its Lagrangian Grassmannian, i.e. the set of Lagrange
subspaces of $E$. Consider the action of the symplectic group
${\rm Sp}(E)$ on the set $\Lambda(E)^3$ of ordered Lagrangian triplets
$(L_1,L_2,L_3)$.
It is clear that the dimensions of the intersections are invariant under this
action. Another independent invariant is the so-called
{\em signature of a Lagrangian triplet} discovered by Hrmander
and Kashiwara \cite{LV80}. Together, these invariants completely
specify the relative position of three Lagrangian subspaces up
to symplectic transformations.
For $(L_1,L_2,L_3)\in\Lambda(E)^3$,
the signature $s(L_1,L_2,L_3)\in {\bf Z}$ is defined as the
signature of the quadratic form
\be Q(L_1,L_2,L_3):\,\, L_1\oplus L_2\oplus L_3\ra {\bf R},\,\,
  (x_1,x_2,x_3)\mapsto \omega(x_1,x_2)+\omega (x_2,x_3)+ \omega (x_3,x_1).\ee
It is immediate from the definition that the signature
$s:\Lambda(E)^3\ra {\bf Z}$ is invariant under symplectic transformations
and antisymmetric under exchange of two of the $L$'s.
Let us list some less trivial properties \cite{GS77,LV80}:
\bprop \label{prop1}

\begin{enumerate}
\item $ s(L_2,L_3,L_4)-s(L_1,L_3,L_4)+s(L_1,L_2,L_4)-s(L_1,L_2,L_3)=0.$
\item {Reduction lemma}: For arbitrary subspaces $K$ of
           $L_1\cap L_2+L_2\cap L_3+
           L_3\cap L_1$,
           \be s(L_1,L_2,L_3)=s(L_1^K,L_2^K,L_3^K),\ee
           where $L_i^K$ denotes the image of $L_i$ under the symplectic
           reduction
           $(K+K^\omega) \ra E^K:=(K+K^\omega)/(K\cap K^\omega)$.

\item
      The signature runs through all integers between $-\f{1}{2}\dim E^F$ and
      $+\f{1}{2}\dim E^F$, where $F=(L_1\cap L_2)+(L_2\cap L_3)+(L_3\cap L_1)$.
      Consequently, $s(L_1,L_2,L_3)+\dim(L_1\cap L_2)+\dim (L_2\cap L_3)
      +\dim(L_3\cap L_1)+n$ is an even number.
\item
      The orbits of the action of ${\rm Sp}(E)$ on $\Lambda(E)^3$ are
      completely
      determined by $\dim(L_1\cap L_2\cap L_3)$, $\dim(L_1\cap L_2)$,
      $\dim(L_2\cap L_3)$,
      $\dim (L_3\cap L_1)$ and $s(L_1,L_2,L_3)$: If these
      five numbers coincide for two triplets, they lie on the same orbit.
\item The signature is locally constant on the set of all triplets with given
   dimensions of intersections.
\end{enumerate}
\eprop
Consider now two continuous paths $L_1,L_2:[a,b]\ra \Lambda(E)$, where
$a\le b$.
Choose a sufficiently fine partition
$a=t_0\le \ldots \le t_k=b$ and Lagrangian subspaces $M_\nu$ such that
$M_\nu$ is transversal to all $L_i(t)$ with
$t_{\nu-1}\le t\le t_\nu$, $i=1,2$.
We define the {\em intersection number}
$[L_1:L_2]$ by
\be [L_1:L_2]=\f{1}{2} \sum_{\nu=1}^k\Big(s(L_1(t_{\nu-1}),L_2
                                     (t_{\nu-1}),M_\nu)
                                      -s(L_1(t_\nu),L_2(t_\nu),M_\nu)\Big).
                                      \label{e.2}\ee
\noindent
Thanks to proposition \ref{prop1}, this definition is indeed independent of
the specific choices.
Since the expressions $s(L_1,L_2,M)$ are locally
constant in $M$ as long as $M\cap L_i=\{0\}$, the above definition also
makes sense
if one considers a symplectic vector bundle over $[a,b]$ and replaces the $L_i$
by Lagrangian subbundles.

\bprop \label{prop2}
 (Properties of the intersection number.)
 \begin{enumerate}
 \item Antisymmetry: $[L_1:L_2]+[L_2:L_1]=0$.
 \item Invariance: $[A(L_1):A(L_2)]=[L_1:L_2]$ for all continuous paths
 $A:[a,b]\ra {\rm Sp(E)}$.
 \item $[L_1:L_2]+\f{1}{2}\dim(L_1(a)\cap L_2(a))+\f{1}{2}\dim(L_1(b)
       \cap L_2(b))
       \in{\bf Z}$.
       In particular, $[L_1:L_2]$ is an integer if the intersections at the
       endpoints are transversal.
 \item
       $[L_1:L_2]+[L_2:L_3]+[L_3:L_1]=\f{1}{2}\Big(s(L_1(a),L_2(a),L_3(a))-
       s(L_1(b),L_2(b),L_3(b))\Big).$
 \item Consider the space of paths $L_1\times L_2:[a,b]\ra \Lambda(E)^2$ with
       given dimensions of the intersections at the endpoints. $[L_1:L_2]$
       labels the connected components of this space.
 \end{enumerate}
\eprop
Using the intersection number, one arrives at a straightforward construction
of the so-called {\em Leray index} \cite{GS77,LV80}.
Let $\pi:\tilde{\Lambda}(E)\ra \Lambda(E)$ denote the universal covering
of the Lagrange-Grassmann manifold. For $u_0,u_1\in\tilde{\Lambda}(E)$, choose
any path $L:[0,1]\ra\Lambda(E)$ such that $L(0)=\pi(u_0)$ and $L(1)=\pi(u_1)$.
Define the Leray index $m(u_0,u_1)\in\f{1}{2}{\bf Z}$ by
\[ m(u_0,u_1)=[L(t):L(1)]=[L(t):L(0)].\]
Proposition \ref{prop2} guarantees that this is independent of the chosen path
and immediately leads to the following statements:

\bprop\label{prop3} (Properties of Leray's index.)
\begin{enumerate}
\item For $L_i=\pi(u_i)$, Leray's formula holds:
\be m(u_1,u_2)+m(u_2,u_3)+m(u_3,u_1)
=\f{1}{2}s(L_1,L_2,L_3).\ee
\item
For arbitrary lifts $u_i(\cdot)$ of Lagrangian curves $L_i(\cdot)$,
\be   [L_1:L_2]=m(u_1(a),u_2(a))-m(u_1(b),u_2(b)).\ee
\item $m(u_1,u_2)$ is locally constant on the set of all $u_1,u_2$ with
fixed $\dim(L_1\cap L_2)$.
\end{enumerate}
\eprop
Let $\tau:\widetilde{\rm Sp}(E)\ra {\rm Sp}(E)$ denote the universal covering
group of the symplectic
group.
Elements $\tilde{A}$ of the covering group can be identified with
homotopy classes of paths $A(t)$ in ${\rm Sp}(E)$ connecting the identity to
$A=\tau(\tilde{A})$.
Recall that the graph
\[ \Gamma_B:=\{(Bx,x)|x\in E\}\]
of a symplectic transformation $B$ in $E$
is a Lagrangian subspace of $E\times E^-$, which is $E\oplus E$ with the
symplectic
form ${\rm pr}_1^*\omega -{\rm pr}_2^*\omega$. We hence obtain an index
\be \mu:\widetilde{\rm Sp}(E)\ra \f{1}{2}{\bf Z},\,\, \tilde{A}\mapsto
[\Delta:\Gamma_{A(t)}],\ee
where $\Delta$ is the graph of the identity, i.e. the diagonal in
$E\times E^-$. (Equivalent indices are introduced in
\cite{CZ84} and
\cite{D76}.)

\bprop\label{prop4}
(Properties of the index $\mu$.)
\begin{enumerate}
\item $\mu(\widetilde{A})$ is locally constant on the set of all $\tilde{A}$
      with
      given $\dim({\rm ker}(A-I))$.
\item $\mu(\tilde{A})+\f{1}{2}\dim({\rm ker}(A-I))\in{\bf Z}$.
\item Let $A(\cdot):[0,1]\ra{\rm Sp}(E)$ be any path representing $\tilde{A}$,
      and let $L,M\in\Lambda(E)$ be arbitrary. Then
      \be\mu(\tilde{A})=[M:A(t)L]+\f{1}{2}s(\Delta,L\times M,\Gamma_A).
      \label{eq&}\ee
      If
      ${\rm ker}(A-I)$ is symplectic and if $L$ is $A$-invariant, the second
      term on the rhs vanishes.
\item (See ref. \cite{CZ84}.) Two elements of the set of all $\tilde{A}$ with
      ${\rm ker}(A-I)=\{0\}$ are in the same connected component if
      and only if they have the same index.

\end{enumerate}
\eprop

\section{The index of periodic trajectories}

Let us now return to the situation described in the introduction. It follows
from the methods in \cite{CZ84} that there is at most one index
${\rm ind}(\gamma,L)$ satisfying properties I1, I2.    On the other
hand, we can give an explicit expression for the index as follows.

Let $q\in \gamma^\sharp$.
Let ${\cal E}_1\subset T_{\gamma^\sharp}Q$ be the tangent bundle of the
orbit cylinder and ${\cal E}_2$ its symplectic orthogonal.
Choosing a symplectic trivialization of these bundles, we can consider the
linearized flow as curves of linear symplectic transformations in
$E_1={\cal E}_1(q)$ and $E_2={\cal E}_2(q)$ respectively. Write
$P(t)$ (Poincare map) for the flow in $E_2$, and choose any Lagrangian
subspace $M$ of $E_1\oplus E_2$. Set
\be {\rm ind}(\gamma,L)=[L(\gamma(t)):M]+\mu\big(\widetilde{P(T)}\big).\ee
It follows from the above propositions that this is well-defined,
in particular it does not depend on the choice of the trivializations.
There is an alternative formula for the index that does not require
any such trivialization. Letting $\Gamma_{TF^T}$ be the graph of the
linearized flow, one has the following formula
formula:
\be
{\rm ind}(\gamma,L)=[L(\gamma(t)):TF^t(M)]+\f{1}{2}s(\Delta,L_q\times M,
\Gamma_{TF^T})+\f{1}{2}{\rm sgn}\left(\f{\partial T}{\partial E}\right).
\ee
where $M\in\Lambda(T_qQ)$ is arbitrary.

\end{document}